\documentclass[letterpaper,prl,twocolumn,reprint,superscriptaddress]{revtex4-1}

\usepackage{color}
\usepackage{graphicx}
\usepackage{dcolumn}
\usepackage{bm}

\begin{document}



\title{Electronically enhanced layer buckling and Au-Au dimerization in epitaxial LaAuSb films}


\author{Patrick J. Strohbeen}
\affiliation{Materials Science and Engineering, University of Wisconsin Madison}

\author{Dongxue Du}
\affiliation{Materials Science and Engineering, University of Wisconsin Madison}

\author{Chenyu Zhang}
\affiliation{Materials Science and Engineering, University of Wisconsin Madison}

\author{Estiaque H. Shourov}
\affiliation{Materials Science and Engineering, University of Wisconsin Madison}

\author{Fanny Rodolakis}
\affiliation{Argonne National Laboratory, 9700 South Cass Avenue, Argonne, Illinois 60439, USA}

\author{Jessica L. McChesney}
\affiliation{Argonne National Laboratory, 9700 South Cass Avenue, Argonne, Illinois 60439, USA}

\author{Paul M. Voyles}
\affiliation{Materials Science and Engineering, University of Wisconsin Madison}

\author{Jason K. Kawasaki}
\email{jkawasaki@wisc.edu}
\affiliation{Materials Science and Engineering, University of Wisconsin Madison}

\date{\today}

\begin{abstract}

We report the molecular beam epitaxial growth, structure, and electronic measurements of single-crystalline LaAuSb films on Al$_2$O$_3$ (0001) substrates. LaAuSb belongs to a broad family of hexagonal $ABC$ intermetallics in which the magnitude and sign of layer buckling have strong effects on properties, e.g., predicted hyperferroelecticity, polar metallicity, and Weyl and Dirac states. Scanning transmission electron microscopy reveals highly buckled planes of Au-Sb atoms, with strong interlayer Au-Au interactions and a doubling of the unit cell. This buckling is four times larger than the buckling observed in other $ABC$s with similar composition, e.g. LaAuGe and LaPtSb. Photoemission spectroscopy measurements and comparison with theory suggest an electronic driving force for the Au-Au dimerization, since LaAuSb, with a 19-electron count, has one more valence electron per formula unit than most stable $ABC$s. Our results suggest that the electron count, in addition to conventional parameters such as epitaxial strain and chemical pressure, provides a powerful means for tuning the layer buckling in ferroic $ABC$s.
\end{abstract}

\pacs{Valid PACS appear here}
\maketitle

\begin{figure}[h]
	\includegraphics[width=\linewidth]{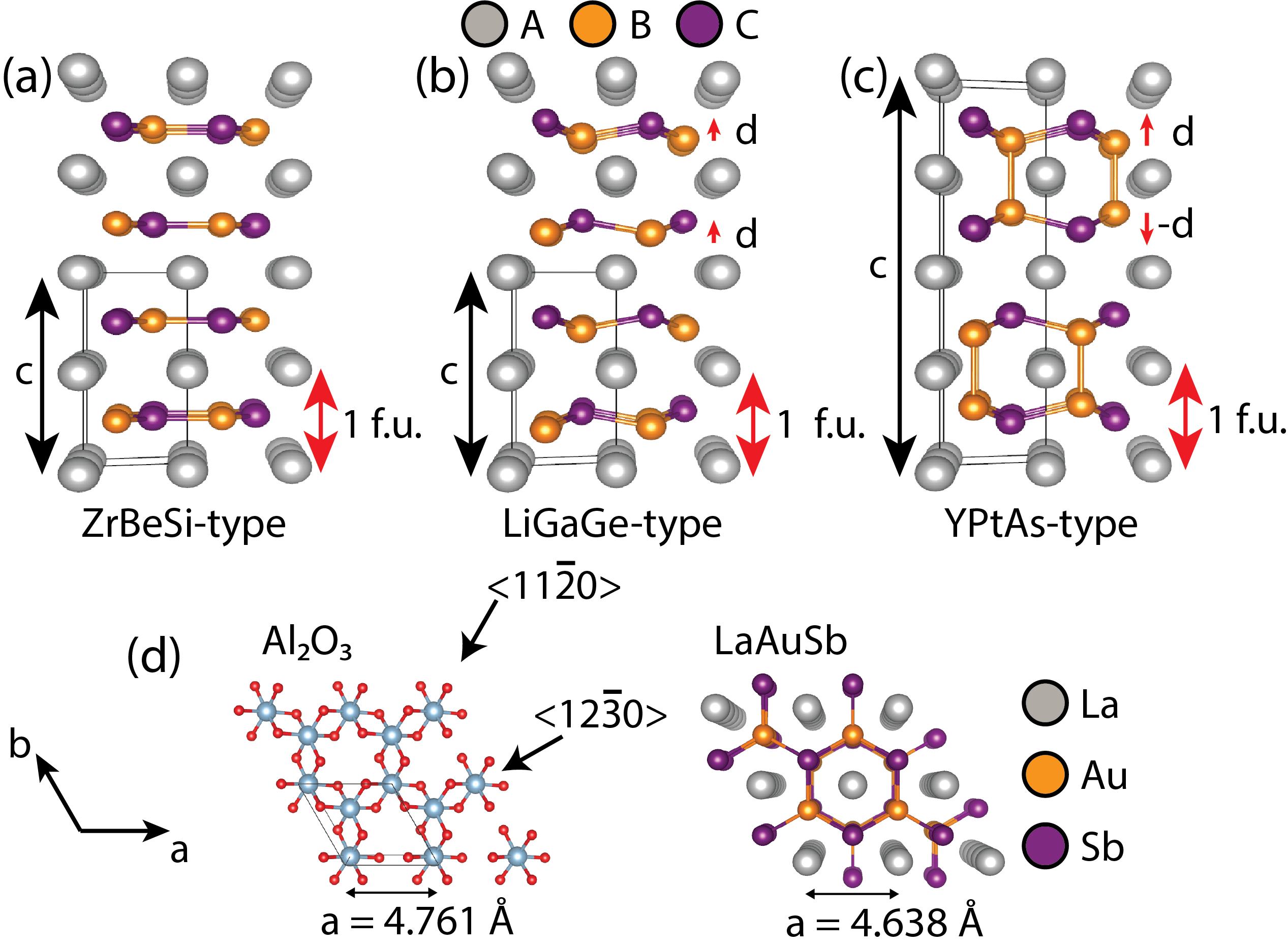}
    \caption{Structural distortions in hexagonal $ABC$ compounds \cite{vesta}. (a) The centrosymmetric ZrBeSi-type structure, consisting of graphite-like $BC$ planes. (b) The polar LiGaGe-type structure, defined by unidirectional buckling \textit{d} of the $BC$ planes. In both ZrBeSi and LiGaGe-type structures, the stacking of the $BC$ planes follows a $B-C-B-C$ sequence. (c) The centrosymmetric YPtAs-type structure consisting of an alternating ``up-down'' buckling of the $BC$ planes and a doubled unit cell along the $c$ axis. The $BC$ planes follow a $B-B-C-C$ stacking sequence. (d) In-plane crystal structures of the $ABC$ compounds and comparison to the sapphire substrate.}
    \label{HexABC}
\end{figure}

Structural distortions, especially layer buckling, are key to understanding and controlling the ferroic properties of hexagonal ternary intermetallics (composition $ABC$) \cite{gao2018,bennett2012hexagonal,narayan2015}. For example, consider the parent centrosymmetric ZrBeSi-type structure (space group $P6_3/mmc$), which consists of planar graphitic $(BC)^{n-}$ layers that are ``stuffed'' with an $A^{n+}$ spacer (Fig. \ref{HexABC}(a)). Compounds with this structure are typically semimetals or Dirac semimetals, e.g., LaCuSn \cite{casper2008} and BaAgBi \cite{du2015}. Upon decreasing the $A^{n+}$ cation size via isovalent substitution, the $(BC)^{n-}$ planes typically buckle in a unidirectional pattern due to increased interlayer interactions, to yield the polar LiGaGe-type structure \cite{bennett2012hexagonal,seibel2015gold} (space group $P6_3mc$, Fig. \ref{HexABC}(b)). Here, the polar distortion gives rise to polar metallicity in the metallic state (e.g. LaAuGe, LaPtSb \cite{ddu2018,benedek2016ferroelectric}) and predicted hyperferroelectricity in the insulating state (e.g. LiZnAs, NaZnSb), in which the long-range polarization is robust against the depolarizing field \cite{bennett2012hexagonal, garrity2014}. In such compounds the properties are determined by both the magnitude and the long-range ordering of the planar buckling, \textit{d}, defined as the displacement along the $c$ axis between dissimilar atoms in the buckled $BC$ plane \cite{bennett2012hexagonal}. Due to this breaking of inversion symmetry, Weyl nodes are also predicted to exist in many LiGaGe-type $ABC$s (e.g. KMgBi, LiZnBi) \cite{disante2016, gao2018, cao2017}. Tuning the magnitude and sign of buckling is proposed to change the crystal momenta and chirality of the Weyl nodes, or cause them to merge into a single Dirac point \cite{gao2018}. All of these predicted properties are \textit{defined} by the buckling \textit{d} in the system, therefore it is crucial to understand the origins of the buckling observed in these structures.\par

Recent experiments and theory suggest that in addition to cation size, the electron count may be an important handle for controlling the layer buckling. Whereas most stable $ABC$s have 18 (or 8) valence electrons per formula unit corresponding to a filled $s^2 p^6 d^{10}$ (or $s^2p^6$) configuration, the family $Ln$AuSb ($Ln = $ lanthanide) has 19 valence electrons. The extra electron destabilizes the unidirectionally buckled LiGaGe-type structure, instead favoring a highly buckled structure with significant Au-Au interlayer bonding (Fig. \ref{HexABC}c, YPtAs-type) \cite{seibel2015gold}. Importantly, although the resulting dimer buckled structure is centrosymmetric, the magnitude of buckling in the AuSb planes is predicted to be much larger than that observed in the LiGaGe-type polar structure \cite{seibel2015gold}. First principles calculations also suggested that LaAuSb hosts a Dirac cone within 100 meV of the Fermi energy \cite{seibel2015gold}. Therefore, an understanding of the coupling between electronic structure (electron count) and layer buckling in $Ln$AuSb may be a path towards tuning the buckling, and hence ferroic properties, of LiGaGe-type compounds.\par

Here we use molecular beam epitaxy (MBE) to demonstrate the first epitaxial growth of LaAuSb. Our large area single crystalline films enable detailed structure and electronic property measurements and provide a path towards integration in layered heterostructures. The films are single crystalline and epitaxial to the $c$-plane Al$_2$O$_3$ substrate, as measured by x-ray diffraction (XRD) and reflection high energy electron diffraction (RHEED). Scanning transmission electron microscopy (STEM) measurements reveal that the AuSb layer buckling is four times larger than the buckling in the 18-valence compounds LaPtSb and LaAuGe. Photoemission spectroscopy measurements of the valence bands are consistent with density functional theory calculations, and suggest that buckling results from an electronic instability that is suppressed in the buckled Au-Au dimer structure. Our results provide a route towards tuning the layer buckling and potential ferroic order in hexagonal $ABC$s.\par

\begin{figure}
	\includegraphics[width=\linewidth]{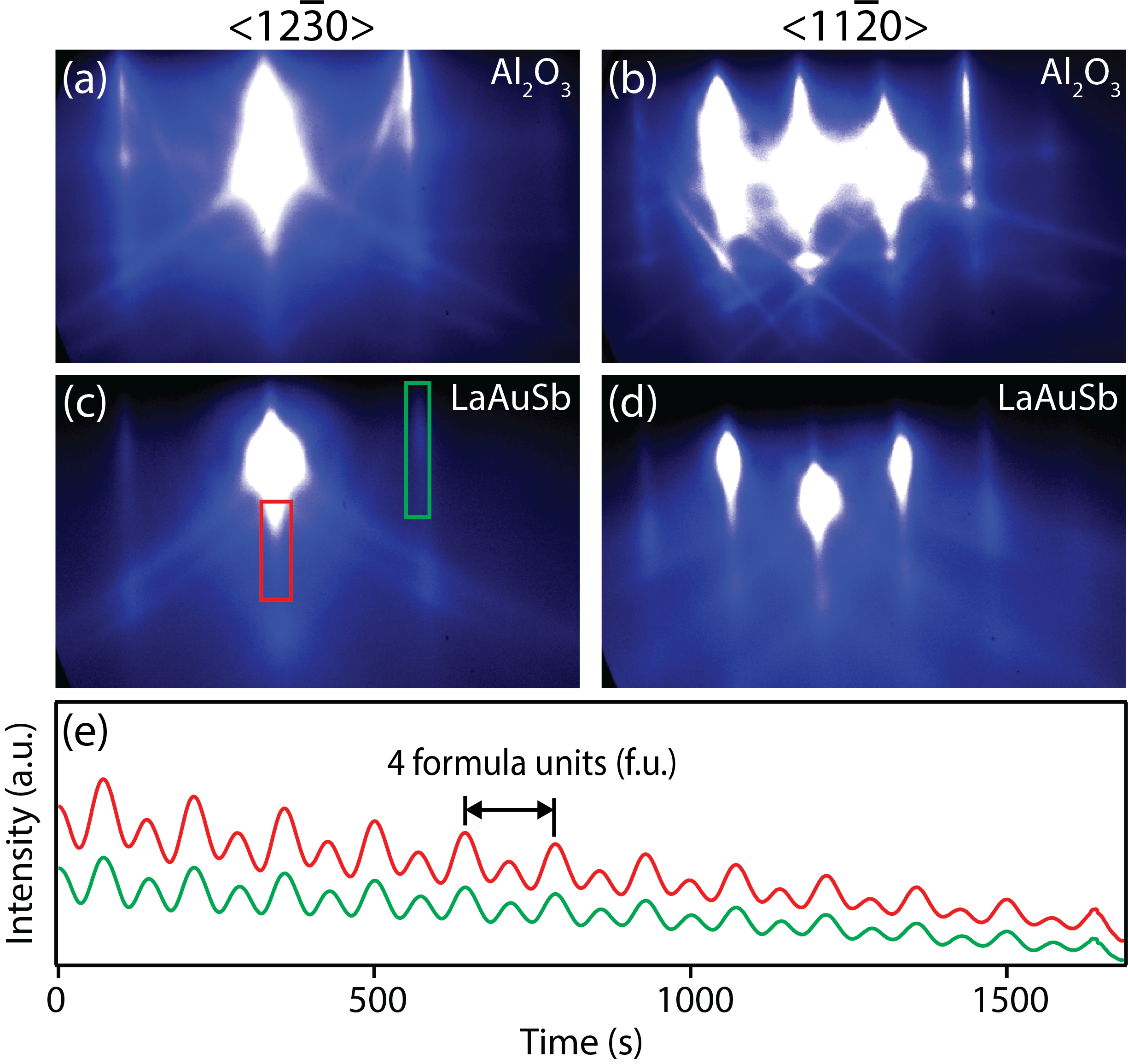} 
    \caption{RHEED patterns and intensity oscillations. (a,b) RHEED patterns of the Al$_2$O$_3$ substrate along the  [12$\overline{3}$0] and 10$\overline{1}$0] azimuths, respectively. (c,d) The corresponding RHEED patterns for the LaAuSb film. (e) RHEED intensity oscillations of the specular (red) and +1 (green) spots observed during growth. The corresponding integration boxes are shown in panel (c).}
    \label{rheed}
\end{figure}

LaAuSb films were grown in a custom MBE system (MANTIS Deposition) on Al$_2$O$_3$ (0001) substrates (MTI) at a growth temperature of $650^\circ$C as measured by pyrometer. The lattice mismatch between LaAuSb and Al$_2$O$_3$ is 3.2\% tensile. Following the film growth, most samples were capped with amorphous Ge to prevent oxidation upon removal from the vacuum system. La and Au fluxes of $1.5\times10^{13}$ atoms/cm$^{2}$s were supplied from effusion cells, as measured by an \textit{in-situ} quartz crystal monitor (QCM). Due to the high relative volatility of Sb, a 30\% excess flux of Sb ($2 \times 10^{13}$ atoms/cm$^2$s) was supplied using a cracker cell, similar to the strategy used for cubic Heusler compounds \cite{patel2014surface, kawasaki2014growth, kawasaki2018simple}. Absolute fluxes were calibrated \textit{ex-situ} by Rutherford Backscattering Spectrometry (RBS).

Figures \ref{rheed}a-d show typical RHEED patterns for the Al$_{2}$O$_{3}$ substrate and the LaAuSb film. The combination of sharp and streaky patterns and Kikuchi lines indicates the relatively smooth epitaxial growth with no apparent secondary phases. We observe persistent RHEED intensity oscillations throughout the entire growth, indicating a layer-by-layer growth mode. A representative example is shown in Fig. \ref{rheed}e, with a fundamental period of 70 s and a doubled beat period of 140 s. By comparison with the expected fluxes from QCM and RBS calibrations, this fundamental period of 70s corresponds to two formula units (f.u.) of LaAuSb (Fig. \ref{HexABC}c).\par

\begin{figure}
	\includegraphics[width=\linewidth]{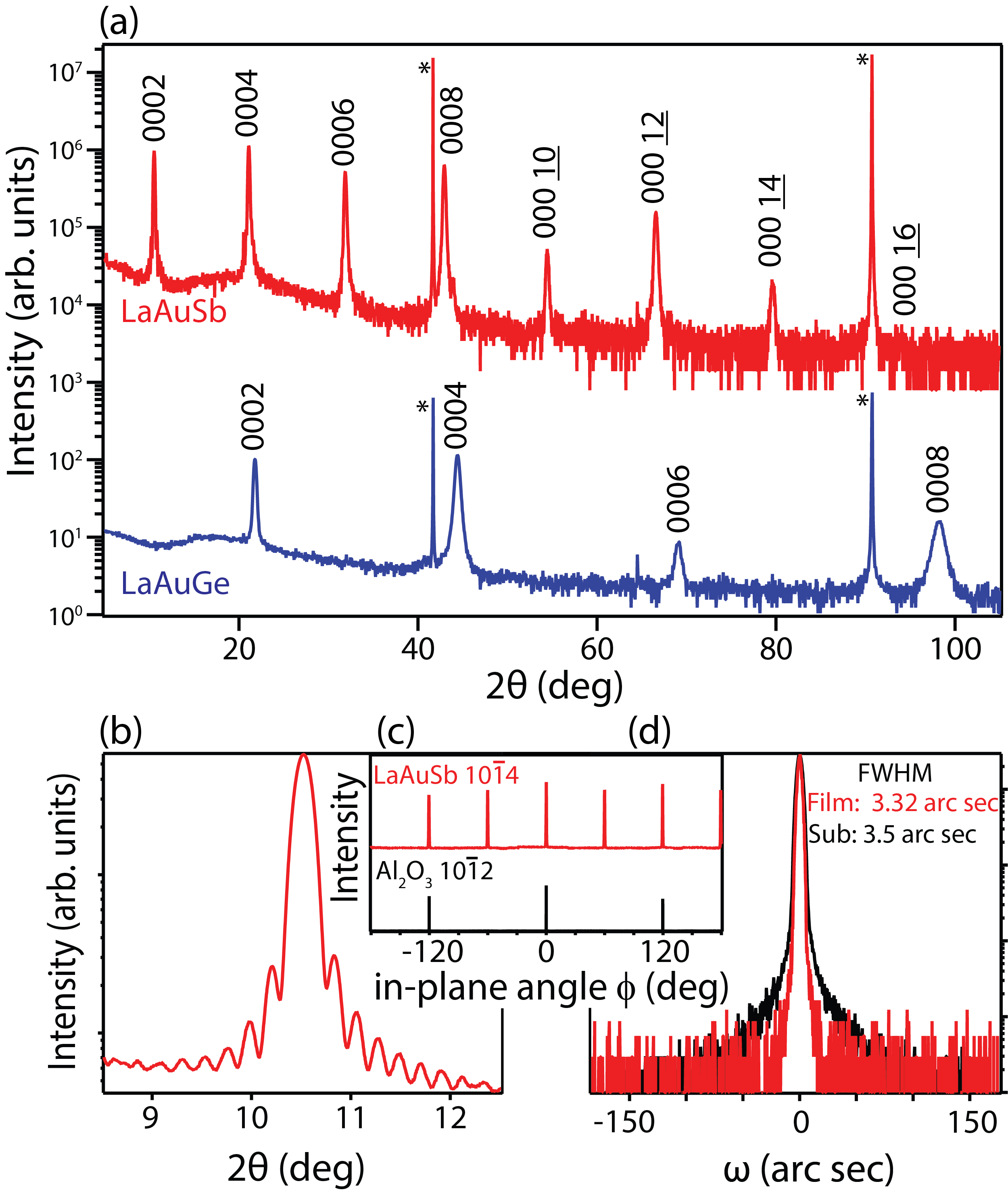}
	\caption{X-ray diffraction of LaAuSb films on Al$_2$O$_3$ (0001). (a) $\theta-2\theta$ scan of LaAuSb film (19 valence electrons) and comparison to LaAuGe (18 valence electrons) \cite{ddu2018}. The LaAuSb shows half-order superstructure reflections, corresponding to a doubled unit cell along the $c$ axis. Reflections from the substrate are marked by asterisks. (b) Higher resolution $\theta-2\theta$ scan around the 0002 reflection of LaAuSb, showing sharp Kiessig fringes. (c) Pole figure of the LaAuSb 10$\overline{1}$4 reflection and sapphire 10$\overline{1}$2 reflection, showing the epitaxial relationship between the film and substrate. (d) Rocking curve scans of the substrate and film showcasing the high crystallinity of this film.}
	\label{xrd}
\end{figure}

Phase purity and single-crystal quality were confirmed \textit{ex-situ} by x-ray diffraction (XRD) using a Panalytical Empyrean diffractometer with Cu $K \alpha$ radiation. All scans used a quadruple-bounce Ge 220 monochromator on the incident beam. A second triple-bounce monochromator inserted in the diffracted beam path was inserted for the rocking curve measurements. Figure \ref{xrd}a shows the $\theta-2\theta$ scan (red online) in which we observe only the expected sharp $000l$ reflections and no secondary phases. For comparison we also plot a 2$\theta$ scan of a LaAuGe film (blue online), which has 18 valence electrons and crystallizes in the undimerized LiGaGe-type structure \cite{ddu2018}. The presence of the 0002, 0006, 000\underline{10}, and 000\underline{14} superstructure reflections in the LaAuSb confirms the doubling of the unit cell along the $c$ axis, consistent with Au-Au dimerization. For samples grown at temperatures below $500^\circ$C, we do not observe the half-order superstructure reflections, suggesting a loss of long-range order. The superstructure-ordered sample grown at $650^\circ$C displays sharp Kiessig fringes (Fig. \ref{xrd}b). The fringe spacing, averaged up to the fifth order, of 0.21 degrees corresponds to a film thickness of 41 nm, or 98 formula units, in good agreement with the 100 formula units expected from RHEED oscillations and QCM fluxes. Rocking curve widths for the film and substrate are sharp and approximately equal (3.32 and 3.5 arc seconds, respectively), indicating a high degree of in-plane ordering (Figure \ref{xrd}d). Pole figure measurements of the film 10$\overline{1}$4 and substrate 10$\overline{1}$2 reflections confirm the $[10\bar{1}0] (0001)_{LaAuSb} \parallel [10\bar{1}0] (0001)_{Al_2 O_3}$ epitaxial relationship (Fig. \ref{xrd}c). From the pole figure and the measured out-of-plane lattice constant of $c=16.79$ \r{A}, we extract an in-plane lattice constant of $a=4.65$ \r{A}. This agrees well with the experimental powder diffraction lattice constants of $c=16.8315$ \r{A} and $a=4.63838$ \r{A} \cite{seibel2015gold}, indicating the 41 nm film on sapphire is nearly fully relaxed to the bulk lattice parameter.\par

\begin{figure}
	\includegraphics[width=\linewidth]{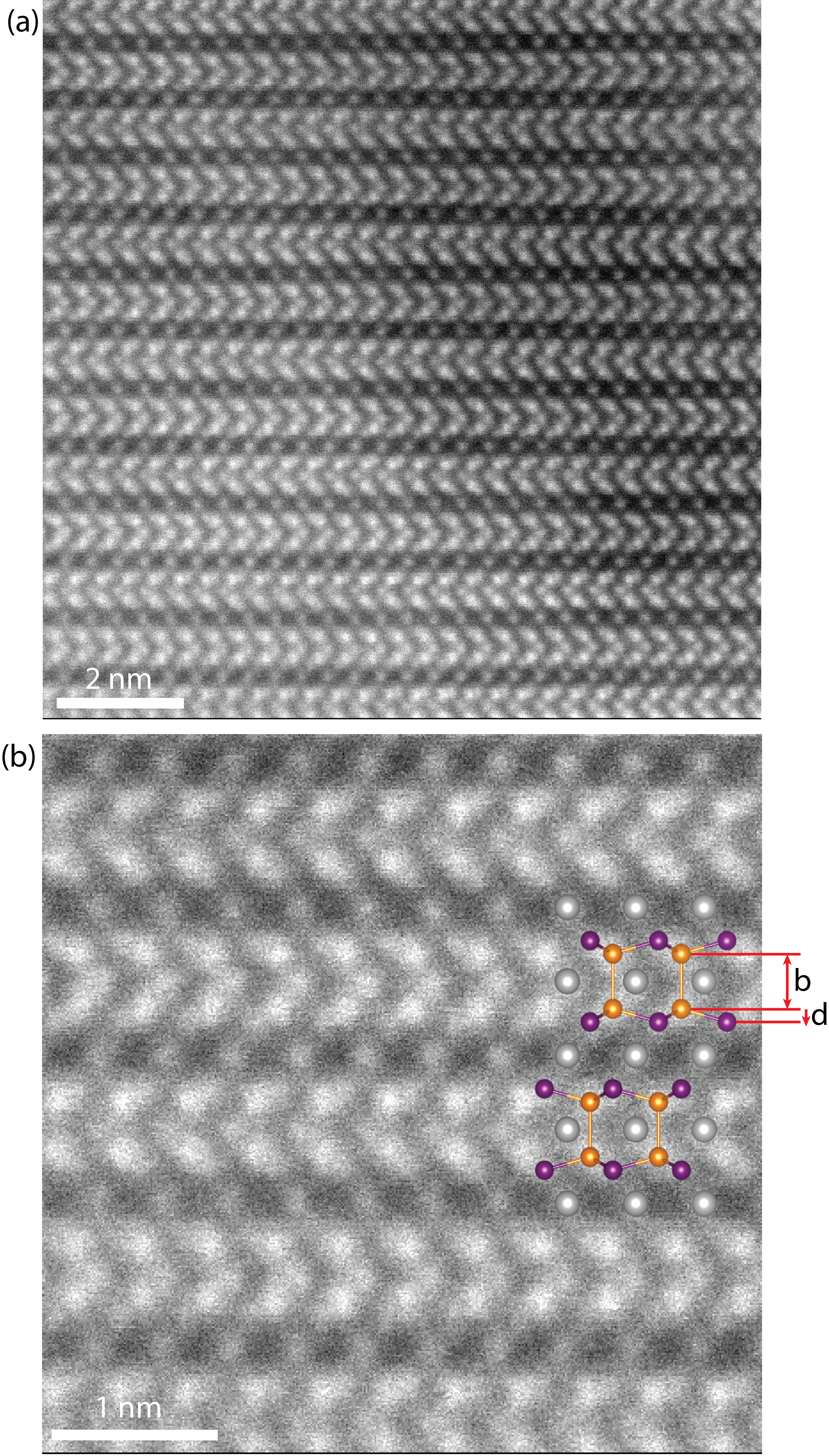}
	\caption{(a) HAADF-STEM image of the LaAuSb film showing the long-range ordering of the Au-Au dimerized structure, oriented along a $[12\overline{3}0]$ zone axis. The $[0001]$ growth direction points upwards. (b) Zoomed in region of the image in (a) to show finer details of the atomic structure. The cartoon ball-and-stick model in (b) is added as a guide to the eye where the grey, gold, and purple balls represent La, Au, and Sb atoms, respectively.}
    \label{tem}
\end{figure}

Scanning transmission electron microscopy (STEM) measurements confirm the Au-Au dimerized structure with a high degree of AuSb layer buckling. These measurements were performed using a FEI Titan STEM equipped with probe corrector using an operating voltage of 200 kV. High angle annular dark field (HAADF) STEM images were collected with a 24.5 mrad probe semi-convergence angle, 18.9 pA probe current, and HAADF detector range of 53.9-269.5 mrad. Annular bright field (ABF) images were collected simultaneously with an ABF detector range of 5.7-12.6 mrad. The sample was prepared for analysis using focused ion beam (FIB) milling. A FIB lamella was lifted out and attached to a Cu support grid prior to the final FIB milling to a thickness of $\approx 100$ nm. The final milling step was done using Ar ion milling using a Fischione Nanomill operated at 900 V, bringing the sample to a final thickness of $\approx 20-40$ nm before being transferred to the TEM column.

The high precision images seen in Figs. \ref{tem}a and b were obtained by applying a non-rigid registration \cite{voylesSTEM} to the HAADF and ABF image series, respectively. Accurate positions of atomic sites were derived by fitting each peak on the HAADF image to a 2D Gaussian function. Sampling across 26 Au-Au pairs, we measure a Au-Au bond length of $b= 3.10 \pm 0.02$ \r{A}, which is in good agreement with the 3.12 \r{A} expected from powder diffraction refinement \cite{seibel2015gold}. Averaging over 51 Au-Sb pairs, we find a planar buckling of $d= 0.80 \pm 0.02$ \r{A} also in good agreement with the 0.75 \r{A} observed in bulk powder diffraction \cite{seibel2015gold}. This buckling is approximately four times larger than the bucklings observed in LiGaGe-type compounds with similar stoichiometry. In comparison, LaAuGe and LaPtSb have \textit{d} of 0.18$\pm$0.01 \r{A} and 0.22$\pm$0.01 \r{A}, respectively \cite{ddu2018}.\par

\begin{figure}
	\includegraphics[width=\linewidth]{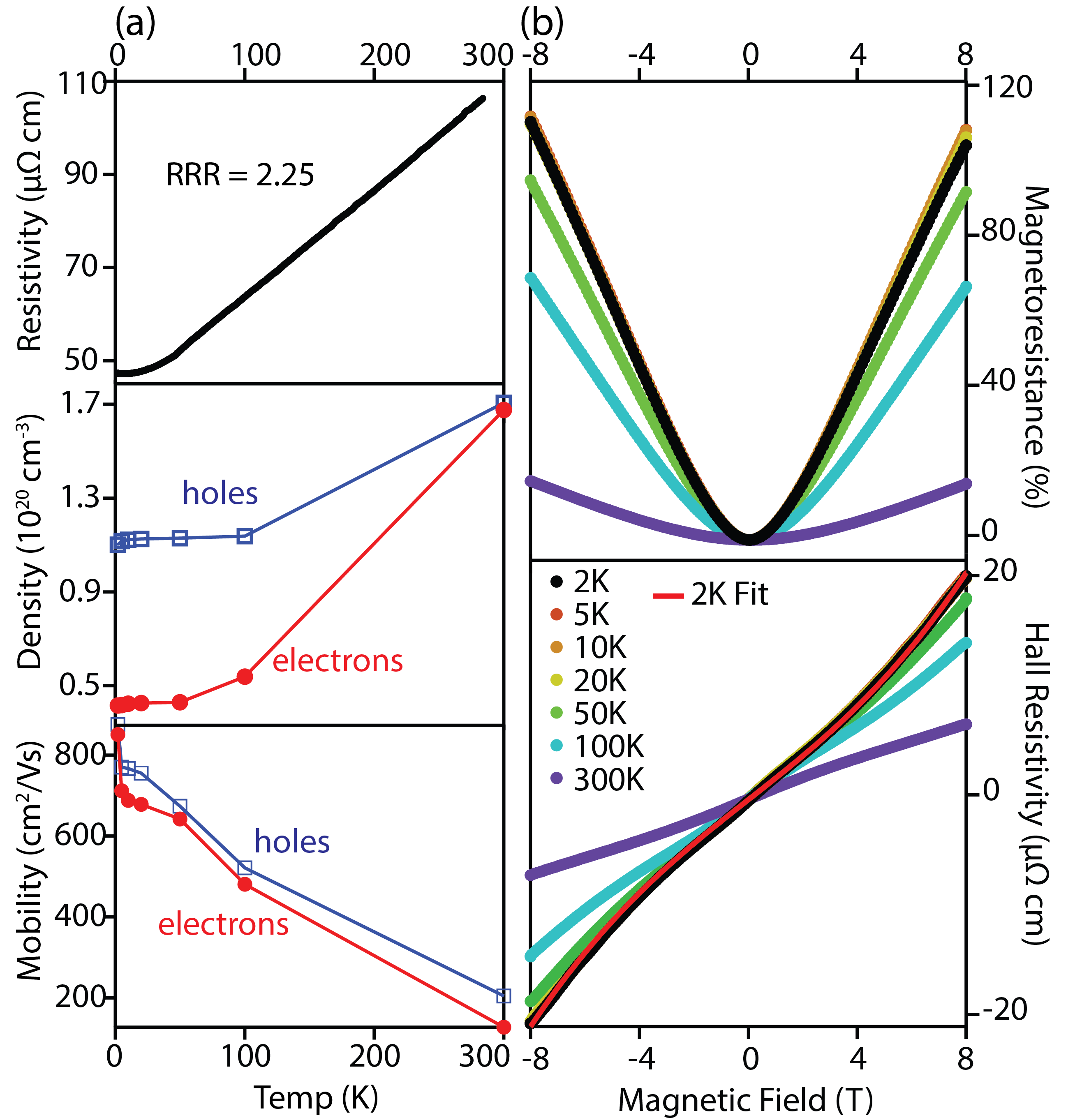}
	\caption{(a) Resistivity (zero-field), carrier densities, and mobilities versus temperature. The densities and mobilities were extracted from a two band fit to the $\rho_{xy}$(B) data. RRR is the residual resistivity ratio ($\rho_{300K}/\rho_{2K}$). (b) (top) Longitudinal magnetoresistance $[\rho_{xx}(B)-\rho_{xx}(0)]/[\rho_{xx}(0)]$. (b) (bottom) Transverse (Hall) resistivity $\rho_{xy}$ versus magnetic field, at select temperatures.}  
	\label{transport}
\end{figure}

Magnetotransport properties were measured using a Quantum Design Physical Property Measurement System (PPMS). The resistivity versus temperature dependence at zero magnetic field shows strong metallic behavior with a residual resistivity ratio (RRR) of $\rho_{300K}/\rho_{2K}$ = 2.25 (Fig. \ref{transport}a (top)). Longitudinal resistivity measurements as a function of out-of-plane magnetic field show magnetoresistance values of up to 109\% at 2K at a field of 8T (Fig. \ref{transport}b (top)). Hall effect measurements show a nonlinearity in $\rho_{xy}$ vs B (Fig. \ref{transport}b (bottom)), which we attribute to multiple carriers since LaAuSb is expected to be a semimetal. Therefore to extract the densities and mobilities we fit to a two-band model of the form \cite{Luo2015}
\begin{equation}
\label{eq1}
\rho_{xy}(B) = \frac{B}{e}\frac{(n_h\mu_h^2-n_e\mu_e^2) + (n_h-n_e)\mu_h^2\mu_e^2B^2}{(n_h\mu_h+n_e\mu_e)^2+(n_h-n_e)^2\mu_h^2\mu_e^2B^2}
\end{equation}
where $n_e$ ($n_h$) and $\mu_e$ ($\mu_h$) are the electron (hole) densities and mobilities, respectively. We first fit the slope of $\rho_{xy}$ vs B in the range $7-8$ T to constrain the difference in carrier densities. We then adjust the concentrations and mobilities to fit $\rho_{xy}$(B) over the full field range, checking that that these parameters are also consistent with the zero-field resistivity $\rho_{xx}(0) = 1/(n_{e}e\mu_e + n_{h}e\mu_h)$.

A representative fit of $\rho_{xy}$ at 2K is shown by the solid red curve in Fig. \ref{transport}b (bottom). The extracted carrier densities and mobilities versus temperature are shown in Fig. \ref{transport}a (middle) and (bottom), respectively. We find weakly varying electron and hole densities of order $10^{20}$ cm$^{-3}$, consistent with expectations for a semimetal, and mobilities approaching 1000 cm$^2$/Vs in the low temperature limit (2K).\par

\begin{figure}
	\includegraphics[width=\linewidth]{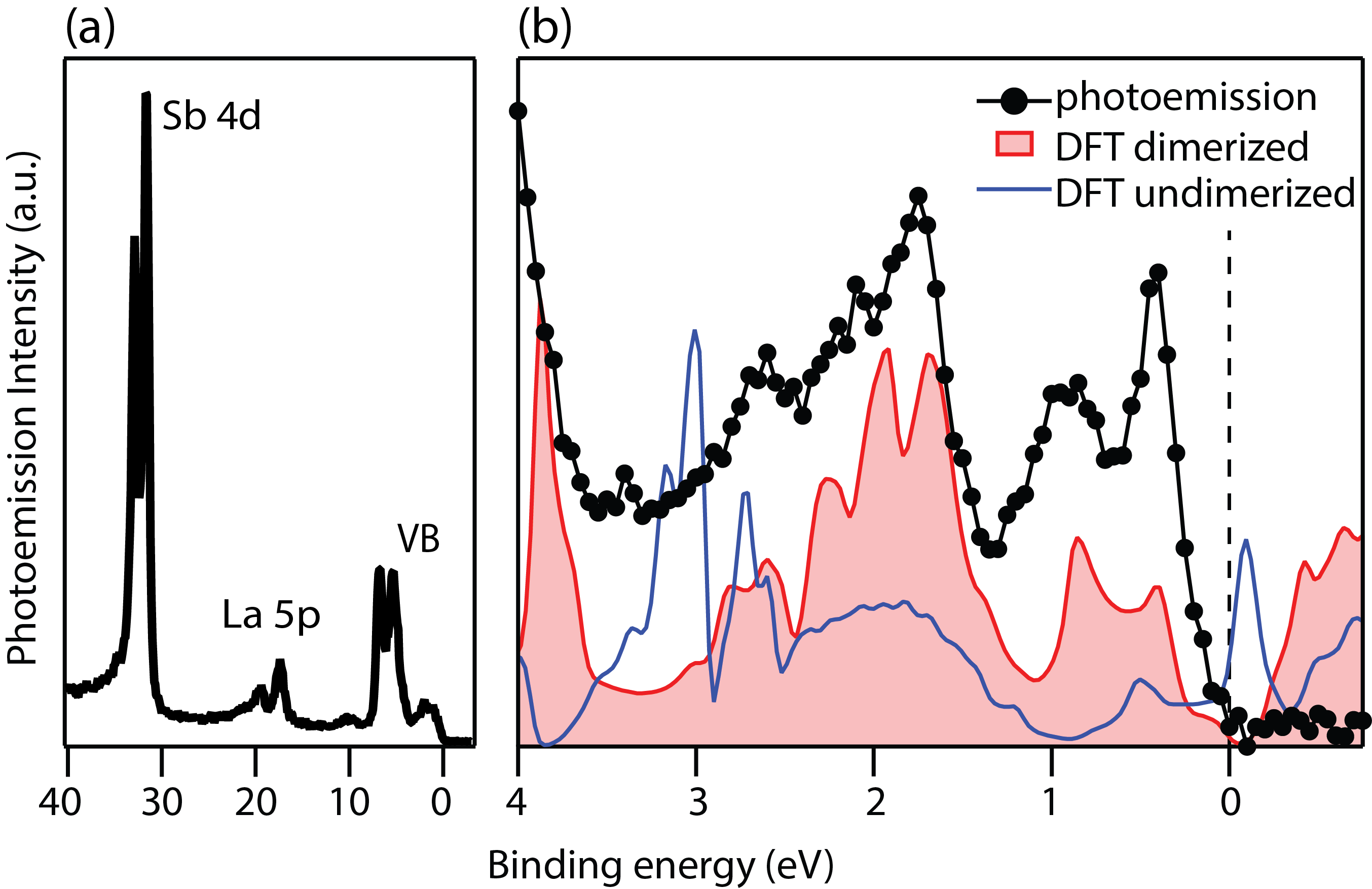}
	\caption{(a) Angle-integrated shallow core levels measured at a photon energy of 1486 eV. (b) Valence band measurement at photon energy of 300 eV and comparison to the DFT (GGA+SO) density of for the dimerized and undimerized structures.}
	\label{photoemission}
\end{figure}

To investigate the origins of the Au dimer buckling and its relation to electronic structure, photoemission spectroscopy measurements were performed at beamline 29-ID of the of the Advanced Photon Source (APS), using a Scienta R4000 analyzer (angular acceptance angle of 14 degrees) and incident photon energies in the range 300 to 2000 eV (Fig. \ref{photoemission}). The Fermi level was referenced measurements of a gold screw that is in electrical contact with the sample. To protect the surfaces, these samples were transported to the APS using an ultrahigh vacuum suitcase (pressure less than $10^{-9}$ Torr, no Ge cap). For comparison with experiment, density functional theory calculations were performed using the Perdew - Becke - Ernzerhof (PBE) parametrization of the generalized gradient approximation including fully relativistic spin-orbit coupling effects (GGA+SO), as implemented in WIEN2k \cite{wien2k}. Further details about the calculation parameters are found in \cite{seibel2015gold}. For the dimerized structure we used the bulk atomic positions as reported in Ref. \cite{seibel2015gold}. The calculated electronic structure is also consistent with Ref. \cite{seibel2015gold}.

We observe good qualitative agreement between the measured angle-integrated valence band spectrum (Fig. \ref{photoemission}, black symbols) and GGA+SO calculation for the dimerized structure (Fig. \ref{photoemission}b, shaded red curve). The measured valence band width is approximately 1 eV and decreases to a sharp but finite minimum at the Fermi energy, consistent with the behavior expected for a semimetal or Dirac semimetal \cite{seibel2015gold}. 

We also compare the photoemission measurement to a GGA+SO calculation for LaAuSb in a hypothetical undimerized LiGaGe-type structure (Fig. \ref{HexABC}b). For this hypothetical structure we fix the in-plane lattice constant and the out-of-plane functional unit spacing to that of the dimerized YPtAs-type structure, and apply the same buckling \textit{d} but in a uniform direction with $B-C-B-C$ layer stacking sequence. 
We find that this structure exhibits a sharp peak in the density of states just above the Fermi energy, which is expected to be unstable (Fig. \ref{photoemission}b, blue curve). In comparison, for the dimerized structure there is a strong suppression in the density of states resulting in a pseudo-gap at the Fermi energy. Our measurements and comparisons with theory suggest an electronic origin for the dimerization, akin to a Peierls distortion, and consistent with the formal electron count of La$_2^{3+}$ (Au-Au)$^0$ Sb$_2^{3-}$ suggested previously by first-principles calculations \cite{seibel2015gold}. In this picture, the Au-Au dimerization is responsible for suppressing the density of states at $E_F$.\par

In summary, we demonstrated the epitaxial single-crystalline growth of the 19 valence electron compound LaAuSb. Due to the strong electronic driving force for Au-Au dimerization, the resultant AuSb layer buckling is four times larger than the buckling observed in 18-electron $ABC$s. Our epitaxial films exhibit relatively large mobility and magnetoresistance with carrier concentrations consistent with that of a compensated semi-metal. Lastly, we showed that our measured valence band structure is consistent with the proposed dimerized electronic structure, providing strong evidence for an electronic driving force towards Au-Au dimerization, and therefore buckling, in these 19 valence electron count compounds.\par

\textit{Acknowledgments.} This work was supported primarily by the United States Army Research Office (ARO Award number W911NF-17-1-0254). Travel for photoemission spectroscopy measurements was supported by the CAREER program of the National Science Foundation (DMR-1752797). This research used resources of the Advanced Photon Source, a U.S. Department of Energy (DOE) Office of Science User Facility operated for the DOE Office of Science by Argonne National Laboratory under Contract No. DE-AC02-06CH11357; additional support by National Science Foundation under Grant no. DMR-0703406. We gratefully acknowledge the use of x-ray diffraction facilities supported by the NSF through the University of Wisconsin Materials Research Science and Engineering Center under Grant No. DMR-1720415. We thank Professor Song Jin for the use of PPMS facilities. We thank Mark Mangus (Eyring Materials Center, Arizona State University) and Greg Haugstad (Characterization Facility, University of Minnesota) for performing RBS measurements.

\bibliographystyle{apsrev}
\bibliography{bibliography}

\end{document}